\documentstyle[12pt]{article}

\font\twlgot =eufm10 scaled \magstep1
\font\egtgot =eufm8
\font\sevgot =eufm7

\font\twlmsb =msbm10 scaled \magstep1
\font\egtmsb =msbm8
\font\sevmsb =msbm7

\newfam\gotfam

\textfont\gotfam\twlgot
\scriptfont\gotfam\egtgot
\scriptscriptfont\gotfam\sevgot

\newfam\msbfam
\textfont\msbfam\twlmsb
\scriptfont\msbfam\egtmsb
\scriptscriptfont\msbfam\sevmsb

\def\pBbb{\relax\ifmmode\expandafter\Bb\else\typeout{You cann't use
Bbb in text mode}\fi}
\def\Bb #1{{\fam\msbfam\relax#1}}

\def\thebibliography#1{\bigskip\section*{\centering References\\}\list
  {\arabic{enumi}.}{\settowidth\labelwidth{#1}\leftmargin\labelwidth
    \advance\leftmargin\labelsep
    \usecounter{enumi}}
    \def\newblock{\hskip .11em plus .33em minus .07em}
    \sloppy\clubpenalty4000\widowpenalty4000
    \sfcode`\.=1000\relax}

\newcommand{\under}[2]{\mathrel{\mathop{#1}\limits_{\scriptstyle #2}}}
\def\op#1{\mathop{\fam0 #1}\limits}

\newcommand{\Id}{{\rm Id\,}}

\newcommand{\ben}{\begin{eqnarray}}
\newcommand{\een}{\end{eqnarray}}
\newcommand{\be}{\begin{eqnarray*}}
\newcommand{\ee}{\end{eqnarray*}}
\newcommand{\bea}{\begin{eqalph}}
\newcommand{\eea}{\end{eqalph}}

\newcommand{\cL}{{\cal L}}

\newcommand{\cH}{{\cal H}}
\newcommand{\cF}{{\cal F}}

\newcommand{\wt}{\widetilde}
\newcommand{\wh}{\widehat}
\newcommand{\ol}{\overline}

\newcommand{\dr}{\partial}

\newcommand{\ip}{\,\rule{1.5mm}{0.2mm}\rule{0.2mm}{4mm}\,\,}

\let\epsilon=\varepsilon
\newcommand{\col}{\colon\ }

\newcounter{eqalph}
\newcounter{equationa}

\newenvironment{eqalph}{\stepcounter{equation}
\setcounter{equationa}{\value{equation}}
\setcounter{equation}{0}

\begin{eqnarray}}{\end{eqnarray}
\setcounter{equation}{\value{equationa}}}

\begin{document}
\hbox{}
\vskip15mm

\centerline {\large\bf Multimomentum Hamiltonian Formalism}
\centerline {\large\bf in Quantum Field Theory}
 \medskip

\centerline {\bf G.Sardanashvily}
\medskip
\centerline {Department of Theoretical Physics,}
\centerline {Moscow State University, 117234, Moscow, Russia}
\bigskip

The familiar generating functionals in quantum field
theory fail to be true measures and, so they make the sense only in the
framework of the perturbation theory. In our approach, generating
functionals are defined strictly as the Fourier transforms of Gaussian
measures in nuclear spaces of multimomentum canonical variables when
 field momenta correspond to derivatives of fields with respect to all
world coordinates, not only to time.

\section{Introduction}

\quad Contemporary field models are almost always the constraint ones. In
order to describe them, one can apply the covariant
multimomentum generalization
of the Hamiltonian formalism in mechanics \cite{6sar,sard,sard10}.
The multimomentum canonical variables are field
functions $\phi^i$ and  momenta $p^\lambda_i$ associated with
derivatives of $\phi^i$ with respect to all world coordinates $x^\mu$,
not only the time.

In classical field theory, if a Lagrangian density is degenerate, the
system of the Euler-Lagrange equations becomes underdetermined and
requires additional conditions. In gauge theory, these are gauge
conditions which single out a representative from each gauge class. In
general case, the above-mentioned supplementary conditions remains
elusive. In the framework of the multimomentum Hamiltonian formalism,
one obtaines them automatically because a part of the Hamilton
equations play the role of gauge conditions. The key point consists in
the fact that, given a degenerate Lagrangian density, one must consider
a family of associated multimomentum Hamiltonian forms in order to
exaust solutions of the Euler-Lagrange equations \cite{6sar,sard}.
There exist the
comprehensive relations between Lagrangian and multimomentum
Hamiltonian formalisms for degenerate quadratic and affine densities.
The most of field models are of these types. As a result, we get the
general procedure of describing constraint systems in classical field theory.

 The present work is devoted to the multimomentum quantum field theory.
This theory as like as the well-known  current algebra
models, has been hampered by the lack of satisfactory commutation
relations between multimomentum canonical variables \cite{car,gun}.
We are based on
the fact that the operation of chronological product of quantum bosonic
fields is commutative and, so Euclidean chronological forms can be
represented by states on commutative tensor alebras. Therefore,
restricting our consideration to generating functionals of Green
functions, we can overcome the difficulties of establishing the
multimomentum
commutation relations.

Moreover, the multimomentum quantum field theory may incorporate
together the canonical and algebraic approaches to quantization of
fields. In physical models,
 the familiar expression
\[
N^{-1}\exp [i\int L(\phi)]\op\prod_x [d\phi(x)]
\]
of a generating functional fails to be a true measure
since the Lebesgue measure
in infinite-dimensional linear spaces is not defined in general.

In the algebraic quantum field theory,  generating functionals of
chronological forms result from the Wick rotation of
the Fourier transforms of Gaussian measures in the duals to nuclear
spaces \cite{cam,S91,S91a}. The problem has consisted in constructing
 such measures. In the present work, we get these measures in terms of
multimomentum canonical variables. They have the universal
form due to the canonical splitting of multimomentum Hamiltonian forms.
In particular, we reproduce the Euclidean propagators of scalar fields
and gauge potentials \cite{sard9}. Note that the covariant
 multimomentum canonical quantization can be
generalized to any field model with a degenerate
quadratic Lagrangian density.

\section{Multimomentum Hamiltonian Formalism}

We consider the
multimomentum generalization of the familiar Hamiltonian formalism to
 fibred manifolds $\pi\col E\to X $ over a $n$-dimensional
base $X$, not only $X={\bf R}$.
 If sections of  $E$ describe classical
fields, one can apply this formalism to field
theory.
In this case, the Legendre manifold
\begin{equation}
 \Pi=\op\wedge^nT^*X\op\otimes_E TX\op\otimes_EV^*E
\label{00}
\end{equation}
plays the role of the finite-dimensional phase space of fields.
 By $VE$ and
$V^*E$ are meant the vertical tangent and cotangent bundles over
 a fibred manifold $E$.
   Given
an atlas of fibred coordinates $(x^\mu, y^i)$ of $E$, the Legendre
manifold is provided with the linear adapted coordinates  $(x^\mu, y^i,
p^\mu_i)$. In this coordinates, a multimomentum Hamiltonian form on
$\Pi$ and the corresponding Hamilton equations read
\ben
&&H=p^\lambda_i dy^i\wedge\omega_\lambda-\cH\omega=
p^\lambda_i dy^i\wedge\omega_\lambda-p^\lambda_i\Gamma^i_\lambda \omega
-\wt\cH\omega,\label{05}\\
&&\dr_\lambda y^i=\dr^i_\lambda\cH, \qquad \dr_\lambda
p^\lambda_i =-\dr_i\cH,\label{06}\\
&&\omega=dx^1\wedge\ldots\wedge dx^n, \qquad  \omega_\lambda
=\dr_\lambda\ip\omega,\nonumber
\een
where $\Gamma$ is a connection on $E$ and $\wt\cH\omega$ is a
horizontal density on $\Pi\to X$.

The multimomentum Hamiltonian formalism is associated with the
Lagrangian formalism in jet manifolds where the jet
manifold $J^1E$ of $E$ plays the role of a finite-dimensional
configuration space.
 The jet manifold $J^1E$ comprises classes
$j^1_x\phi$ of sections $\phi$ of $E$ which are identified by the first
two terms of their Taylor series at points $x$.
 It is provided
with the adapted coordinates $(x^\lambda, y^i, y^i_\lambda)$ where
\[
y^i_\lambda(j^1_x\phi)=\dr_\lambda \phi^i(x).
\]

A first-order Lagrangian
density $L=\cL (x^\lambda, y^i, y^i_\lambda)\omega$
on $J^1E$ defines the Legendre
morphism $\wh L$ of $J^1E$ to $\Pi$:
\[
(x^\lambda,y^i,p^\lambda_i)\circ\wh L=(x^\lambda,y^i,\pi^\lambda_i),
\qquad \pi^\lambda_i=\dr^\lambda_i\cL.
\]
Conversely, a multimomentum Hamiltonian form $H$ on $\Pi$ defines
the momentum morphism $\wh H$ of $\Pi$ to $J^1E$:
\[
(x^\lambda,y^i,y^i_\lambda)\circ\wh H=(x^\lambda,y^i,\dr^i_\lambda \cH).
\]
We say that a multimomentum Hamiltonian form $H$ is associated with a
Lagrangian density $L$ if
\be
&&
\wh L\circ\wh H|_Q=\Id Q, \qquad Q=\wh L(J^1E),\\
&&\cL(x^\mu, y^i, \dr^i_\mu\cH)=p^\lambda_i\dr^i_\lambda\cH-\cH.
\ee
In general, different multimomentum Hamiltonian forms may be associated
with the same Lagrangian density. The most of field models meet
 the following relations between
Lagrangian and multimomentum Hamiltonian formalisms.

(i)
All multimomentum
Hamiltonian forms $H$ associated with a Lagrangian density
$L$ are equal to each other on the constraint space $Q$, that is,
$H|_Q=H_L$. Moreover, for every section $\phi$ of $E$, we have
\begin{equation}
(\wh L\circ j^1\phi)^*(H_L)=(j^1\phi)^*(L)=L(\phi). \label{025}
\end{equation}

(ii)
 If a solution $r$ of the Hamilton equations (\ref{06})
corresponding to a multimomentum Hamiltonian form $H$ associated with a
Lagrangian
density $L$ belongs to the constraint space $Q$, then $\pi_\Pi\circ
r$ is a solution
of the Euler-Lagrange equations for $L$. Conversely, for each
local solution $s$ of the Euler-Lagrange equations defined by a
Lagrangian density
$L$, there exists an associated multimomentum Hamiltonian form
such that $\wh L\circ s$ is a solution of the corresponding Hamilton
equations.

The relation (\ref{025}) gives the reason to use sections $r$
of the Legendre manifold $\Pi\to X$ as functional variables
in quantum field theory. On physical level, one
can consider the naive generating functional
\ben
&&Z=N^{-1}\int\exp [i\int (r^*H+\alpha_i\phi^i\omega+
\alpha^i_\mu p^\mu_i\omega)]\op\prod_x [d\phi^i(x)][dp^\mu_i(x)],
 \label{026}\\
&&r^*H=(p^\mu_i(x)\dr_\mu \phi^i(x)-\cH)\omega. \nonumber
\een
Note that the canonical splitting (\ref{05}) of multimomentum
Hamiltonian forms leads to standard terms $p^\mu_i\dr_\mu\phi^i$ in
generating functionals in multimomentum canonical variables.

The generating functional (\ref{026}) fails to be a true measure. The problem
of representation of generating functionals by measures
can be settled in the framework of the algebraic
quantum field theory \cite{S91,S91a}.

\section{Algebraic Quantum Field Theory}

In accordance with the algebraic approach, a quantum field system can
be characterized by a topological $^*$-algebra $A$
 and by a continuous  state $f$ on $A$.
To describe particles, one considers usually a tensor algebra $A_\Phi$
of  a real linear locally convex topological space $\Phi$ endowed
with involution operation.
We further assume that $\Phi$ is a nuclear space.

In the axiomatic quantum field theory of real scalar fields,
the quantum field algebra is $A_\Phi$  with $\Phi={\bf R}S_4$. By ${\bf
R}S_m$ is meant the real subspace of the nuclear
Schwartz space $S({\bf R}^m)$
of complex functions $\phi(x)$ on ${\bf R}^m$ such that
\[
\|\phi\|_{k,l}=\op{\max\ \ \sup}_{|\alpha|\leq l, x\in{\bf R}^m}
(1+|x|)^k
\frac{\dr^{|\alpha|}}{(\dr x^1)^{\alpha_1}
\ldots(\dr x^m)^{\alpha_m}}\phi (x),
\]
\[
|\alpha|=\alpha_1+\cdots+\alpha_m,
\]
is finite for any collection $(\alpha_1,\ldots,\alpha_m)$ and all
$l,k\in{\bf Z}^{\geq 0}$. A state $f$ on $A_{{\bf R}S_4}$
is represented by a
family of temperate distributions $W_n\in S'({\bf R}^{4n})$:
\[
f(\phi_1\cdots\phi_n)=
\int W_n(x_1,\ldots,x_n)\phi_1(x_1)\cdots\phi_n(x_n)
d^4x_1\ldots d^4x_n.
\]
If $f$ obeys the
Wightman axioms, $W_n$ are the familiar $n$-point Wightman functions.

To describe particles created at some moment and
destructed at another moment, one uses the chronological forms $f^c$
given by the expressions
\begin{equation}
W^c_n(x_1,\ldots,x_n)=
\op\sum_{(i_1\ldots i_n)}\theta(x^0_{i_1}-x^0_{i_2})
\cdots\theta(x^0_{i_{n-1}}-x^0_{i_n})W_n(x_1,\ldots,x_n) \label{030}
\end{equation}
where $(i_1\ldots i_n)$ is a rearrangement of numbers $1,\ldots,n$.
The forms (\ref{030}) fail to be distributions and do
not define a state on $A_{{\bf R}S_4}$. At the same time, they
issue from the Wick rotation of the
Euclidean states on $A_{{\bf R}S_4}$ describing
particles in interaction zone.

Since chronological forms (\ref{030}) are symmetric,
Euclidean
forms can be introduced as states on a commutative tensor algebra.
Note that they differ
>from the Swinger functions associated with the Wightman functions.

Let $B_\Phi$ be the commutative quotient of $A_\Phi$. This
algebra can be regarded as the enveloping algebra of the Lie algebra
associated with the Lie commutative group $G_\Phi$ of translations in
$\Phi$. We therefore can construct a state on the algebra
$B_\Phi$ as a vector form
of its cyclic representation induced by a strong-continuous unitary
cyclic representation of $G_\Phi$. Such representation is characterized
by a positive-type continuous generating function $Z$ on $\Phi$,
that is,
\[
Z(\phi_i-\phi_j)\alpha^i\ol\alpha^j\geq 0, \qquad Z(0)=1,
\]
for all collections of $\phi_1,\ldots,\phi_n$ and
 complex numbers $\alpha^1,\ldots,\alpha^n$. If the
function $\alpha\to Z(\alpha\phi)$ on ${\bf R}$ is analytic at 0 for
each $\phi\in \Phi$, the positive continuous form $F$ on $B_\Phi$ is
given by the expressions
\[
F_n(\phi_1\cdots\phi_n)=i^{-n}\frac{\dr}{\dr\alpha^1}
\cdots\frac{\dr}{\dr\alpha^n}Z(\alpha^i\phi_i)|_{\alpha^i=0}.
\]

In virtue of the well-known theorem \cite{gel}, any
function $Z$ of the above-mentioned type is the Fourier
transform of a positive bounded measure $\mu$
in the dual $\Phi'$ to $\Phi$:
\begin{equation}
Z(\phi)=\op\int_{\Phi'}\exp[i
\langle w,\phi\rangle]d\mu(w) \label{031}
\end{equation}
where $\langle,\rangle$ denotes the contraction between $\Phi'$ and
$\Phi$. The corresponding representation of $G_\Phi$ is given by operators
\[
g(\phi)\col u(w)\rightarrow\exp[i\langle w,\phi\rangle]u(w)
\]
in the space of quadraticly $\mu$-integrable functions $u(w)$ on
$\Phi'$, and we have
\[
F_n(\phi_1\cdots\phi_n)=\int\langle w,\phi_1\rangle\cdots\langle w,\phi_n
\rangle d\mu(w).
\]

For instance, a generating function $Z$ of a Gaussian state $F$ on $B_\Phi$
 reads
\begin{equation}
Z(\phi)=\exp[-\frac12 \Lambda(\phi,\phi)] \label{032}
\end{equation}
where the covariance form $\Lambda(\phi_1,\phi_2)$ is a positive-definite
Hermitian bilinear form on $\Phi$, continuous in $\phi_1$ and $\phi_2$.
This generating function is the Fourier transform of a
Gaussian measure in $\Phi'$. The forms $F_{n>2}$ obey the Wick rules where
\[
F_1=0, \qquad F_2(\phi_1,\phi_2)=\Lambda(\phi_1,\phi_2).
\]
In particular, if $\Phi={\bf R}S_n$, the covariance form of a Gaussian
state is uniquelly defined by a distribution $W\in S'({\bf R}^{2n})$:
\[
\Lambda(\phi_1,\phi_2)=\int W(x_1,x_2)\phi_1(x_1)\phi_2(x_2)
d^nx_1d^nx_2.
\]

In field models, a generating function $Z$ plays the role of a
generating functional represented by the functional integral
(\ref{031}). If $Z$ is the Gaussian generating function (\ref{032}),
its covariance form $\Lambda$ defines Euclidean
propagators. Propagators of fields on the Mincowski space are
reconstructed by the Wick rotation of $\Lambda$ \cite{sard10}.

\section{Scalar Fields}

Let $E$ be a vector bundle over a world manifold $X^4$.
Its sections describe
scalar matter fields. In jet terms, their Lagrangian density reads
\be
&&L_{(m)}=\frac{1}{2}a^E_{ij}[g^{\mu\nu}(y^i_\mu-\Gamma^i_\mu)
(y^j_\nu-\Gamma^j_\nu)-m^2y^iy^j]\sqrt{|g|}\,\omega,
\\
&&\Gamma^i_\mu =\Gamma{}_m{}^i{}_j(x)y^j, \qquad g=\det\,g_{\mu\nu},
\ee
where $a^E$ is a fibre metric in $E$, $\Gamma$ is a linear connection
on $E$ and $g$ is a world metric on $X^4$. Because of the canonical
vertical splitting $VE=E\times E$, the corresponding Legendre manifold
 (\ref{00})  is
\[
\op\wedge^nT^*X\op\otimes_E TX\op\otimes_E E^*.
\]
The Legendre morphism $\wh
L_{(m)}$ and the unique multimomentum
Hamiltonian form associated with the Lagrangian density $L_{(m)}$
are given by the expressions
\ben
&&p^\lambda_i\circ\wh
L_{(m)}=g^{\lambda\mu}a^E_{ij}(y^j_\mu-\Gamma^j_\mu), \nonumber\\
&&H_{(m)}=p^\lambda_idy^i\wedge\omega_\lambda-p^\lambda_i
\Gamma^i_\lambda\omega-
\frac12(a^{ij}_Eg_{\mu\nu}p^\mu_ip^\nu_j|g|^{-1}+\nonumber\\
&& \qquad m^2a^E_{ij}y^iy^j)\sqrt{|g|}\,\omega\label{027}
\een
where $a_E$ is the fibre metric in $E^*$ dual to $a^E$.

For the sake of
simplicity, we here examine scalar fields without symmetries
Let $\wt\phi$ be real Euclidean scalar fields on the Euclidean space
$X={\bf R}^4$. The corresponding Legendre manifold
\[
\wt\Pi=(\op\wedge^4T^*X\op\otimes_XTX)\op\times_X{\bf R}
\]
is provided with the
adapted coordinates $(z^\mu, \wt y, \wt p^\mu)$. Sections
$r$ of $\wt\Pi$
are represented by functions $(\wt\phi(z), \wt p^\mu(z))$
on ${\bf R}^4$ which take their values in the vector space
\[
V=(\op\wedge^4{\bf R}_4\otimes{\bf R}^4)\times{\bf R}, \qquad {\bf
R}_4=({\bf R}^4)'.
\]
Their commutative tensor algebra is $B_\Phi$ where $\Phi=V\otimes
{\bf R}S_4$. The scalar form
\[
\langle r|r\rangle_\Phi=\int[\delta_{\mu\nu}\wt p^\mu(z)\wt p^\nu(z)
+\wt\phi^2(z)]d^4z
\]
brings $\Phi$ into the rigged Hilbert space. Let
\[
H_{(m)}
=\wt p^\mu d\wt y\wedge\omega_\mu-\frac12(-\delta_{\mu\nu}\wt p^\mu
\wt p^\nu+m^2\wt y^2)
\omega
\]
be the multimomentum Hamiltonian form describing Euclidean scalar
fields. The covariance form $\Lambda$ of the associated generating
function is defined by the relation
\begin{equation}
\int 2r^*H_{(m)}
=\langle r|\gamma r\rangle_\Phi=-\Lambda(\gamma r,\gamma
r), \qquad r\in\Phi, \label{034}
\end{equation}
where $\gamma$ is a first order linear differential operator on $\Phi$.
We have
\ben
&&\Lambda(r,r)
=\int[\Delta_F(z_1,z_2)\wt\phi(z_1)\wt\phi(z_2)+\wt p^\mu(z_1)
\frac{\dr\Delta_F}{\dr z^\mu_1}\wt\phi(z_2)+\nonumber\\
&&\wt\phi(z_1)\frac{\dr\Delta_F}{\dr z^\mu_2}\wt p^\mu(z_2)
+(-\delta_{\mu\nu}
\delta^{(4)}(z_1-z_2)+\frac{\dr^2\Delta_F}
{\dr z^\mu_1\dr z^\nu_2})\wt p^\mu(z_1)
\wt p^\nu(z_2)]d^4z_1d^4z_2,\nonumber\\
&&\Delta_F(z_1,z_2)=\int\Delta_F(q)\exp[iq(z_1-z_2)]d_4q,\nonumber\\
&& \Delta_F(q)=(m^2+\delta^{\mu\nu}q_\mu q_\nu)^{-1},\label{M3}
\een
where $\Delta_F$ is the Feynman propagator of Euclidean scalar fields.

\section{Gauge Theory}

\quad Let $P\to X^4$ be a principal bundle with a structure Lie
group $G$ of internal symmetries.
There is the 1:1~correspondence between principal connections on $P$ and
global sections $A^C$ of the affine bundle
$
C=J^1P/G
$
modelled on the vector bundle
\begin{equation}
\ol C =T^*X \otimes V^GP,\qquad V^GP=VP/G.\label{042}
\end{equation}
The bundle $C$ is provided with  the fibred coordinates
$(x^\mu,k^m_\mu)$ such that its
section $A^C$ has the coordinate expression
\[
(k^m_\mu\circ A^C)(x)=A^m_\mu(x)
\]
where $A^m_\mu(x)$ are coefficients of a local connection
1-form. In gauge theory, sections $A^C$ are treated as
gauge potentials.

The configuration space of gauge potentials is the jet manifold $J^1C$.
It is provided with the adapted coordinates $(x^\mu, k^m_\mu,
k^m_{\mu\lambda})$.
There exists the canonical splitting
\ben
&& J^1C=C_+\under{\oplus}{C}C_-=(J^2P/G)\under{\oplus}{C}
(\stackrel{2}{\wedge}T^*X\otimes V^GP),\nonumber\\
&& (k^m_\mu,s^m_{\mu\lambda},\cF^m_{\lambda\mu})=
(k^m_\mu,k^m_{\mu\lambda}+
k^m_{\lambda\mu} + c^m_{nl}k^n_\lambda k^l_\mu,
k^m_{\mu\lambda}-
k^m_{\lambda\mu} -c^m_{nl}k^n_\lambda k^l_\mu) \label{38}
\een
where $c^m_{nl}$ are the structure constants of the Lie algebra
${\bf g}$ of the group $G$.  In the coordinates~(\ref{38}),
the conventional
Yang-Mills Lagrangian density $L_{(A)}$ of gauge potentials
is given by the expression
\begin{equation}
L_{(A)}=\frac14 a^G_{mn}g^{\lambda\mu}g^{\beta\nu}\cF^m_{\lambda
\beta}\cF^n_{\mu\nu}\sqrt{|g|}\,\omega \label{40}
\end{equation}
where $a^G$ is a $G$-invariant metric in the Lie algebra ${\bf g}$ and
$g$ is a world metric on $X^4$.

For gauge potentials, we have the Legendre manifold
\[
\Pi=
\stackrel{4}{\wedge}T^*X\otimes TX\under{\otimes}{C}[C\times\ol C]^*
\]
provided with the canonical coordinates $(x^\mu, k^m_\mu,
p^{\mu\lambda}_m)$. This is a phase space of gauge potentials. It also
has the canonical splitting
\[
p^{\mu\lambda}_m=p^{(\mu\lambda)}_m +  p^{[\mu\lambda]}_m
=\frac12(p^{\mu\lambda}_m+
p^{\lambda\mu}_m) + \frac12(p^{\mu\lambda}_m-
p^{\lambda\mu}_m).
\]
The fibred
manifold $\Pi\to X$ is the affine bundle modelled on the vector
bundle
\begin{equation}
\op\wedge^4T^*X\op\otimes_{\ol C}V^*\ol C\to X. \label{043}
\end{equation}

The Legendre
morphism corresponding to the Lagrangian density (\ref{40}) is
\bea
&&p^{(\mu\lambda)}_m\circ\wh L_{(A)}=0, \label{65a}\\
&&p^{[\mu\lambda]}_m\circ
\wh L_{(A)}=a^G_{mn}g^{\lambda\alpha}g^{\mu\beta}
\cF^n_{\alpha\beta}\sqrt{|g|}. \label{65b}
\eea
The multimomentum Hamiltonian forms associated with the
Lagrangian density (\ref{40}) read
\ben
&&H_B=p^{\mu\lambda}_mdk^m_\mu\wedge\omega_\lambda-
p^{\mu\lambda}_m
\ol\Gamma^m_{\mu\lambda}\omega-\wt\cH\omega, \nonumber\\
&&\wt\cH= \frac14 a^{mn}_Gg_{\mu\nu}
g_{\lambda\beta} p^{[\mu\lambda]}_m p^{[\nu\beta]}_n{|g|}^{-1/2},
\nonumber\\
&&\ol\Gamma^m_{\mu\lambda}=\frac{1}{2} [c^m_{nl}k^n_\lambda k^l_\mu
+\dr_\mu B^m_\lambda+\dr_\lambda B^m_\mu-c^m_{nl}
(k^n_\mu B^l_\lambda+k^n_\lambda B^l_\mu)]-\nonumber\\
&& \qquad \Gamma^\beta_{\mu\lambda}(B^m_\beta-k^m_\beta), \label{66}
\een
where $B$ is some section of $C$,
 $\ol\Gamma$ is a connection on $C$ and $\Gamma^\beta_{\mu\lambda}$
are Christoffel symbols of a world metric $g$. We have
\[
H_B|_Q=p^{[\mu\lambda]}_mdk^m_\mu\wedge\omega_\lambda- \frac12
p^{[\mu\lambda]}_m c^m_{nl} k^n_\lambda k^l_\mu\omega-\wt\cH\omega.
\]

The Hamilton equations corresponding to the
multimomentum Hamiltonian form~(\ref{66}) read
\ben
&&\dr_\lambda p^{\mu\lambda}_m=-c^n_{lm}k^l_\nu
p^{[\mu\nu]}_n+c^n_{ml}B^l_\nu p^{(\mu\nu)}_n
-\Gamma^\mu_{\lambda\nu}p^{(\lambda\nu)}_m,
\label{67a}\\
&&\dr_\lambda k^m_\mu+ \dr_\mu
k^m_\lambda=2\ol\Gamma^m_{(\mu\lambda)}\label{67c}
\een
plus the equation
(\ref{65b}). On the constraint space (\ref{65a}), the equations
 (\ref{65b}) and (\ref{67a})
are the familiar Yang-Mills equations.
The equation~(\ref{67c}) plays the role of gauge condition.

 In the
algebraic quantum field theory, only fields forming a linear space
are quantized. We therefore fix a background gauge
potential $B$ and consider deviation fields $\Omega=A^C-B$
which are sections of the vector bundle (\ref{042}).
The corresponding Legendre bundle (\ref{043}) is endowed with the adapted
coordinates
\begin{equation}
(x^\mu,\ol
k^m_\mu,p^{\mu\lambda}_m)=(x^\mu,k^m_\mu-B^m_\mu(x),p^{\mu\lambda}_m).
\label{M2}
\end{equation}
Sections $r$ of the Legendre bundle (\ref{043}) over the Euclidean
space ${\bf R}^4$ are represented by functions $(\Omega^m_\mu(z),
p^{\mu\lambda}_m(z))$ taking their values in the vector space
\[
F=(\op\wedge^4{\bf R}_4\otimes{\bf R}^8\otimes{\bf g}^*)\times({\bf
R}_4\otimes{\bf g}).
\]
The commutative tensor algebra of $r$
is $B_\Phi$ where $\Phi=F\otimes{\bf R}S_4$. The nuclear space $\Phi$ is
provided with the corresponding scalar form $\langle|\rangle_\Phi$
which brings $\Phi$ into the rigged Hilbert space.

To define a Gaussian state on this algebra, let us consider
 the multimomentum Hamiltonian form   (\ref{66}). In the coordinates
(\ref{M2}), it reads
\ben
&&\ol H_B=p^{\mu\lambda}_md\ol k^m_\mu\wedge\omega_\lambda-
p^{\mu\lambda}_m\ol\Gamma^m_{\mu\lambda}
\omega-\wt\cH_B\omega, \nonumber\\
&&\wt\cH_B= \frac14 a^{mn}_Gg_{\mu\nu}
g_{\lambda\beta} p^{[\mu\lambda]}_m p^{[\nu\beta]}_n{|g|}^{-1/2} +
\frac12 p^{[\mu\lambda]}_m(\cF_B{}_{\mu\lambda}^m+c^m_{nl}\ol
k^n_\lambda \ol k^l_\mu),
\nonumber\\
&&\ol\Gamma_B{}^m_{\mu\lambda}=c^m_{nl}B^n_\lambda\ol k^l_\mu
+\Gamma^\beta_{\mu\lambda}\ol k^m_\beta, \label{044}
\een
where $\cF_B$ is the strength of the background gauge potential $B$, and
$\Gamma_B$ is a connection on $\ol C$ associated with the principal
connection $B$.  One can use the
multimomentum Hamiltonian form (\ref{044}) in order to quantize the
deviation fields $\Omega$ on $X={\bf R}^4$.

For the sake of simplicity, let us assume
that the structure group $G$ is
compact and simple $(a_G^{mn}=-2\delta^{mn})$.
 We have
\ben
&&r^*H_B=r^*H_1+r^*H_2=[\frac12\delta_{\mu\nu}\delta_{\lambda\beta}
\delta^{mn}p^{[\mu\lambda]}_mp^{[\nu\beta]}_n+
p^{\mu\lambda}_m\nabla_\lambda
\Omega^m_\mu]d^4z-\nonumber\\
&&\qquad
\frac12 p^{[\mu\lambda]}_m(\cF_B{}^m_{\mu\lambda}+c^m_{nl}\ol
k^n_\lambda \ol k^l_\mu)d^4z \label{045}
\een
where $\nabla$ denotes the covariant derivative corresponding to the
principal connection $B$.

To construct the associated Gaussian state, we use the quadratic part
$r^*H_1$ of the form (\ref{045}). The term $r^*H_2$ describes
interaction considered by the pertubation theory.

The scalar form $\int r^*H_1$ on $\Phi$ however is degenerate. There
are two ways for this difficulty to be overcomed.

(i) In accordance with the conventional quantization scheme, we can
restrict ourselves to sections $r$ taking values in the constraint
space (\ref{65a}). Since the form
\ben
&&r^*\ol H_B=\frac12[\delta_{\mu\nu}\delta_{\lambda\beta}
\delta^{mn}p^{[\mu\lambda]}_mp^{[\nu\beta]}_n+p^{[\mu\lambda]}_m
\cF^m_{\mu\lambda}]d^4z=\nonumber\\
&&\qquad r^*H_0-\frac12 p^{[\mu\lambda]}_mc^m_{nl}
(\Omega^n_\lambda+B^n_\lambda)(\Omega^l_\mu+B^l_\mu)d^4z\label{046}
\een
is degenerate,
we must then consider the gauge orbit space $\Xi_G$ which is
the quotient of the space $\Xi$ of connections $A^C$ by the group of
gauge transformations. There exists a neighborhood $N$ centered at the
image of $B$ in $\Xi_G$ such that there is a local section $s_B\col
N\to \Xi$ whose
values are elements $A^C\in\Xi$ satisfying the gauge condition
\[
\delta^{\mu\nu}\nabla_\mu(A^C_\nu-B_\nu)=0
\]
\cite{mit}. Hence, $N$ is locally isomorphic to a
Hilbert space. Not discussing here the Gribov
ambiguity problem,
let us assume that there exists a connection $B$ such that
$s_B$ is a global section. If the bilinear part $r^*H_0$ of the form
(\ref{046}) induces a nuclear scalar form on $\Xi'_G$, one can
construct the associated Gaussian measure $\mu$ in $\Xi_G$.
 If $s_B$ is $\mu$-measurable morphism,
there exists a measure $\mu_{GF}$
in $\Xi$ which is the image of $\mu$ with respect to $s_B$. If this
measure exists,
it is concentrated in $s_B(\Xi_G)\subset\Xi$. We call it the
gauge-fixing measure. In contrast with the naive expressions used in
the gauge models, it is not the measure whose base is a Gaussian
measure and density is the Faddeev-Popov determinant. Determinant
densities
are attributes of Lebesgue measures which fail to be defined in
general case.

(ii) The first procedure fails to be applied to  general case of
degenerate field systems, without gauge invariance.
 At the same time, one can insert additional terms quadratic in
$p^{(\mu\nu)}_m$ into the multimomentum Hamiltonian form (\ref{044})
which bring $\int r^*H_1$ into a nondegenerate scalar form. In general
case, we have
\begin{equation}
\ol H'_B=\ol H_B-h\omega, \quad h=\frac12[a_1
\delta_{\mu\nu}\delta_{\lambda\beta}
\delta^{mn}p^{(\mu\lambda)}_mp^{(\nu\beta)}_n+
a_2(p^{\mu\mu}_n)^2],\label{047}
\end{equation}
where $a_1\neq 0$ and $a_2$ are some constants. The Lagrangian
$L'_{(A)}$ associated with the multimomentum Hamiltonian form
(\ref{047}) includes additional terms quadratic in $\ol k^m_{(\mu\nu)}
+ c^m_{nl}k^n_{(\mu}B^l_{\nu)}$.
If the quadratic form $\int r^*H'_1$ is nondegenerate, one can use the
relation similar the relation (\ref{034}) in order to construct the
covariance form $\Lambda_B$ of the associated Gaussian generating
function $Z_B$.

For instance, if $a_1=1$ and $a_2=0$, we have
\[
r^*H'_1=[\frac12\delta_{\mu\nu}\delta_{\lambda\beta}
\delta^{mn}p^{\mu\lambda}_mp^{\nu\beta}_n+p^{\mu\nu}_m\nabla_\lambda
\Omega^m_\mu]d^4z.
\]
The associated covariance form $\Lambda_B$ exists. In particular, it
defines the propagator of Euclidean deviation fields $\Omega$ which
coinciders with the Green's operator of the covariant Laplasian
$\delta^{\mu\nu}\nabla_\mu\nabla_\nu$. This Green's operator exists
\cite{mit}. In the case of $B=0$, the propagator of
deviation fields $\Omega$ coinciders with the familiar propagator of
gauge potentials which corresponds to the Feynman gauge ($\alpha=1$),
but there are no ghost fields. In comparison with the measure
$\mu_{GF}$, the Gaussian measure $\mu_B$ defined by the generating function
$Z_B$ is not concentrated in a gauge-fixing subset.

Note that, after gauge transformations $B\to B'$, measures $\mu_B$ and
$\mu_{B'}$ are not equivalent in general. It means that a gauge phase
of a background gauge potential may be valid, otherwise
electromagnetic potentials.
In the case of an abelian structure group $G$, the
multimomentum Hamiltonian form (\ref{046}), the associated covariance
form $\Lambda_B$ and the Gaussian measure $\mu_B$ are independent on a
background potential $B$.


\begin{thebibliography}{ederf}

\bibitem{cam} D. Campodonico, Hadronic Journal, {\bf 10}, 25 (1987).

\bibitem{car} J. Cari\~nena, M. Crampin,  and L. Ibort,  Diff. Geom. and
 Appl., {\bf 1}, 345 (1991).

\bibitem{gel} I. Gelfand and V. Vilenkin,. {\it Generalized Functions},
Vol.IV (Academic Press, New York, 1964).

\bibitem{gun} C. G\"unter,  J. Diff. Geom., {\bf 25}, 23 (1987).

\bibitem{mit} P. Mitter and C. Viallet, Comm. Math. Phys., {\bf 79}, 457
(1981)

\bibitem{S91} G. Sardanashvily, Nuovo Cimento, {\bf 104A}, 105 (1991).

\bibitem{S91a} G. Sardanashvily and O. Zakharov,  Rep. Math.
Phys. {\bf 29},  101 (1991).

\bibitem{6sar} G. Sardanashvily and O. Zakharov, Diff. Geom. and Appl.
{\bf 3} 245 (1993).

\bibitem{sard} G. Sardanashvily, {\it Gauge Theory in Jet Manifolds}
(Hadronic Press, Palm Harbor, 1993).

\bibitem{sard9} G. Sardanashvily, Hadronic J. {\bf 17} 227 (1994).

\bibitem{sard10} G. Sardanashvily, Multimometum Hamiltonian Formalism,
LaTeX Preprint hep-th/9403172.

\end{thebibliography}
\end{document}